\documentstyle[aps,prl,epsf,twocolumn]{revtex}
\begin{document}
\def\a{\alpha} 
\def\b{\beta}
\twocolumn[\hsize\textwidth\columnwidth\hsize\csname
@twocolumnfalse\endcsname 
\draft
\title{Hidden symmetries in deformed microwave resonators}
\author{Joseph Samuel and Abhishek Dhar}
\address{ Raman Research Institute,
Bangalore 560080, India\\ }
\maketitle
\widetext
\begin{abstract} 
We explain the ``Hidden symmetries'' observed in
wavefunctions of deformed microwave resonators in recent experiments.
We also predict that other such symmetries can be seen in microwave
resonators. 
\end{abstract}

\pacs{PACS numbers: 03.65.-w, 41.20.Bt}]
\narrowtext

Lauber et al \cite{lauber} experimentally studied the pattern of
Berry phases that emerges when a  microwave cavity is
cyclically deformed around a rectangular shape. 
Standing electromagnetic waves in the cavity can be mapped and the
``wave functions'' followed through the cyclic deformation to
measure the Berry phase.
Apart from the Berry phases, which were
primarily of interest in ref\cite{lauber}, those authors also noticed
a curious symmetry: the standing wave patterns at different deformations
are related.
Subsequent theoretical work \cite{Mano,Pistol} has clarified
the pattern of Berry phases seen in the experiment. However, the
``hidden symmetry'' has not been explained so far.
The purpose of this brief report is to provide an understanding of the
``hidden symmetry'' and thus a complete and correct interpretation
of the experiment described in \cite{lauber}.

Consider a rectangular cavity (see Fig.1) with sides $(a,b)$ having $n$ 
degenerate 
modes: the scalar Laplacian $-\nabla^2$ has $n$ degenerate eigenfunctions.
If the cavity is deformed, the degeneracy will in general be broken. 
Let us suppose
that the deformation consists (as in the experiment of ref. \cite{lauber})
of moving the corner around its undeformed position so that the
rectangle is deformed to a quadrilateral.
\vbox{
\vspace{2.0cm}
\epsfxsize=7.0cm
\epsfysize=4.5cm
\epsffile{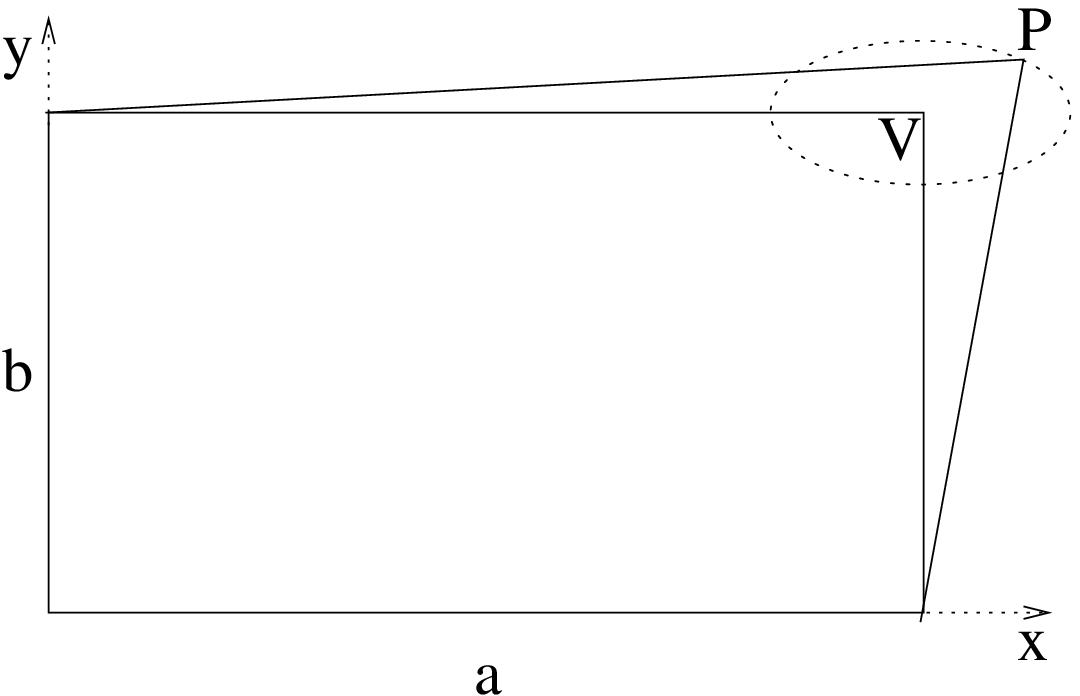}
\begin{figure}
\caption{ A deformation of a rectangle into a quadrilateral. The vertex
$V=(a,b)$ is moved to the point $P=V+(\delta x, \delta y)=V+ab
(\a,\b) $. We consider an experiment where  $P$ is moved around the
elliptic path shown in the figure.     
\label{vplane} 
}
\end{figure}} 
This deformation can be effected in the formalism by performing a
co-ordinate transformation $x=u(1+\alpha v),
y=v(1+\beta u)$, (where $(\alpha,\beta)$ are the deformation parameters) 
which maps the
deformed  rectangle in the $(x,y)$ plane to an undeformed rectangle
in the $(u,v)$ plane. Transforming the Laplacian to curvilinear
$(u,v)$ coordinates, we find  
$H =-\nabla^2= \frac{-1}{\sqrt{g}} \frac{\partial}{\partial x^{\mu}}
\sqrt{g}
g^{\mu\nu} \frac{\partial}{\partial x^{\nu}}$. 
Matrix elements of
$H$ have the form $<\psi_1\vert H \vert \psi_2> =- \int d^2x
\sqrt{g}\psi_{1}^{*}  \nabla^2 \psi_2
 = -\int d^2x \psi_{1}^{*} 
\frac{\partial}{\partial x^\mu} \sqrt{g}g^{\mu\nu} \frac{\partial}{\partial x^\nu} 
\psi_2$. Expanding to first order in $\alpha,\beta$, we then get  $H = H_0 + H_1$, where 
$H_0 = -(\partial_u \partial_u + \partial_v \partial_v)$ and
$H_1 = \alpha f + \beta g$, with $f = vX + uY$, and $g = - uX + v Y$,
expressed in terms of the differential operators $X =
\partial_u 
\partial_u - \partial_v \partial_v$ and $Y = 2\partial_u \partial_v$.

The unperturbed Hamiltonian $H_0$ has the discrete symmetries 
$P_1:u \to a-u,\,\, P_2 : v \to b-v$, the mirror planes of the rectangular
box. We now restrict attention
to the $n$ dimensional degenerate subspace ${\cal H}_n$ of $H_0$ and  
choose eigenstates
of $H_0$ to have definite parity with respect to both these reflections.
In fact, we choose these in the form 
$\vert i > = \vert n_im_i > = 
(2/\sqrt{ab})
\sin \frac{n_iu\pi}{a} \sin \frac{m_i v\pi}{b}$, where
$n_i, m_i$ are positive integers. 
Since the states are all degenerate
eigenstates of  $H_0$, we have $\frac{n_{i}^{2}}{a^2} +
\frac{m_{i}^{2}}{b^2} =
 \frac{n_{j}^{2}}{a^2} + \frac{m_{j}^{2}}{b^2}$ for all $i, j$. In particular
$n_i = n_j \Rightarrow m_i = m_j$. These states are also eigenstates of
$X$ with eigenvalues $\lambda_i = (\frac{n_{i}^{2}\pi^2}{a^2} - 
\frac{m_{i}^{2}\pi^2}{b^2})$. It follows that $<i\vert vX \vert j> = 
\lambda_j <i\vert v\vert j> = \lambda_j <n_i\vert n_j> <m_i\vert v
\vert m_j>= \lambda_i \delta_{ij} <m_i\vert v\vert m_i>$.
From $P_2 vP_2 = (b-v)$, it follows that $< m_i\vert v \vert m_i>=< 
m_i\vert P_2 vP_2 \vert m_i> =
b<m_i\vert m_i> - <m_i\vert v\vert m_i>$. So we conclude that $<m_i\vert 
v \vert m_i> =b/2$ and so, in
${\cal H}_n$,  $vX = bX/2$ and similarly that $u X = aX/2 $. The form
of the perturbations is thus $f = bX/2 + uY,~ g = -aX/2 + vY$. 

The ``mirror symmetry'' observed by Lauber et al in their experiment is
related to the way the unperturbed levels transform under parity. We
consider all possible cases and thus find the necessary and sufficient
conditions for this symmetry to be observed. Let us introduce 
$\sigma_{1i}$ as the $P_1$ parity of the ith state 
($P_1|i>=\sigma_{1i}|i>$) and similarly $\sigma_{2i}$ as the $P_2$ parity
of the ith state. The different cases are listed below with 
an example (for $n=3$) illustrating each non trivial case:
\begin{enumerate}

\item $\sigma_{1i}=\sigma$ and $\sigma_{2i}=\sigma'$ 
for all $i=1,2,...n$ where $\sigma,~\sigma'$ can take 
values $\pm 1$ [{\bf
 Example:} $a=\sqrt{3},~b=1$ and  levels $(2, 6),~(8, 4),~(10,2)$]. 
In this case $<i|u Y |j>=<i|P_2 (P_2 u Y P_2) P_2 |j>=-<i|u Y|j>=0$
 and similarly $<i|v Y |j>=<i|P_1 (P_1 v Y P_1) P_1 |j>=-<i|v Y| 
 j>=0$. Thus $f=bX/2$ and $g=-aX/2$ and this is an uninteresting case
because the perturbations do not span a two dimensional space.   
\item  The product $\sigma_{1i} \sigma_{2i}=\sigma$ for 
all $i$, but $\sigma_{1i}$ and
     $\sigma_{2i}$ are not
     the same for all $i$ [{\bf Example:} $a=\sqrt{3},~b=1$ and  levels
     $(1,3),~(4,2),~(5,1)$].  In this case $<i|u Y |j>=<i|P_2 P_1 (P_1 
     P_2 u Y P_2 P_1) P_1 P_2 |j>=<i|(a-u) Y| j>$ which implies
     $uY=aY/2$. Also $<i|v Y |j>=<i|P_2 P_1 (P_1 P_2 v Y P_2 P_1) P_1
     P_2 |j>=<i|(b-v) Y j>$ and this gives $vY=bY/2$. Thus in this
     case $f=bX/2+aY/2$ and $g=-aX/2+bY/2$. Defining new coordinates:
     $\a=b \a ' + a \b';~ \b=-a \a'+b \b'$ we have $H_1=\a f+ \b g= \a' (bf-a
     g) + \b' (a f+b g) = (a^2+b^2) (\a' X/2+ \b' Y/2)$. Since $P X
     P=X,~P YP=-Y$ 
     for $P=P_1,P_2$, hence we see that wavefunctions at points
     $p(\a',\b')$ and  $p'(\a',-\b')$ can be related either by $P_1$ or
     $P_2$. Proof: Let $H_p |\psi_p>=e
     |\psi_p>$. Then $P_1 H_{1p} P_1 P_1 |\psi_p>=e P_1 |\psi_p>$ or
     $H_{1p'} P_1 |\psi_p >= e P_1 |\psi_p >$ which implies, assuming
     all degeneracies have been lifted, that $|\psi_{p'}>= \pm P_1
     |\psi_p>$. This is the case studied by Lauber et al \cite{lauber}. 
     Note that the $\b'$ axis is 
     along the long diagonal of the rectangular cavity.  

\item  
$\sigma_{1i}=\sigma$ for all $i$, but $\sigma_{2i} $ is not
     the same for all $i$ [{\bf Example:} $a=2,~b=1$ and levels
     $(2,18),~(12,17),~(20,15)$].  In this case $f=b X/2+u Y$ and
     $g=-aX/2$. The coordinate transformation  $\a= a \b';~
     \b= \a'+b \b'$ gives $H_1=\a f+ \b g= \a'  g + \b'
     ( a f+b g) = -\a' aX/2+ \b'au Y$. Since $P_2 X P_2=X$ and $P_2 uY
     P_2=-uY$,  it follows that wavefunctions at points $p(\a',\b')$ and
     $p'(\a',-\b')$  can be related by $P_2$.

\item $ \sigma_{2i}=\sigma$ for all $i$, but $\sigma_{1i} $ is not
     the same for all $i$. This case is similar to (3).

\item Neither of $\sigma_{1i},\sigma_{2i},\sigma_{1i}\sigma_{2i}$ 
is the same
for all $i$. It can be proved that this case cannot be
realized for any choice of $a,b,n_i,m_i$. 
Proof: We enumerate all the  possibilities. We can have
$n_i^2/a^2+m_i^2/b^2= n_j^2/a^2 +m_j^2/b^2$ only if $b^2/a^2$ is
rational. Let $b^2/a^2=p/q$, where $p$ and $q$ are relatively prime. 
We find that $n_i^2p+m_i^2q=n_j^2p+m_j^2q=N$ for all $i,j$. Thus we need to 
consider the following  cases classifed according to the parity (odd or 
even) of 
$(p,q)$ (a) $(o,o)$ 
 (b) $ (o,e)$, 
(c) $(e,o)$,
where $o$ and $e$
denote odd and even parities 
respectively. 
For case (a): if $N$ is even then the states can have
parities $(P_1,P_2)$ either $(-,-)$ or $(+,+)$. If $N$ is odd then
they can have parity 
$(+,-)$ or $(-,+)$. Thus the only combinations we can get  belong to
type (1) or (2). For case (b): if $N$ is even then the states can have
parity $(+,+)$ or $(+,-)$. If $N$ is odd then they can have parity
$(-,+)$ or $(-,-)$.  In this case the possible combinations belong to
type (3). The case (c) leads to type (3).

Thus there are no examples of type (5). 
\end{enumerate}
In summary,
we have explained the mirror symmetry of \cite{lauber} in the framework
of first order perturbation theory (see \cite{Pistol,letter} for
the limitations of this theory) and noticed other situations where such
symmetry may be observed.


\begin{references}
\vspace{-1.5cm}
\bibitem{lauber} H.-M. Lauber, P. Weidenhammer and D. Dubbers,
Phys. Rev. Lett {\bf 72}, 1004 (1994). 
\bibitem{Mano} D.E. Manolopoulos and M.S. Child, Phys. Rev. Lett {\bf 82}, 2223 (1999).
\bibitem{Pistol} F. Pistolesi and N. Manini, Phys. Rev. Lett. {\bf
85}, 1585 (2000).
\bibitem{letter} J. Samuel and A. Dhar, Phys. Rev. Lett. {\bf
87}, 260401 (2001).  
\end{references}
\end{document}